# Measurements and modeling of absorption by $CO_2$+$H_2O$ mixtures in the spectral region beyond the $CO_2$ $\nu_3$-band head

H. Tran[1,*], M. Turbet[1], P. Chelin[2], X. Landsheere[2]

[1]*Laboratoire de Météorologie Dynamique, IPSL, UPMC Univ Paris 06, Ecole polytechnique, Ecole normale supérieure, Sorbonne Universités, Université Paris-Saclay, PSL Research University, CNRS, 4 place Jussieu, 75005, Paris, France*

[2]*Laboratoire Interuniversitaire des Systèmes Atmosphériques (LISA, CNRS UMR 7583). Université Paris Est Créteil, Université Paris Diderot, Institut Pierre-Simon Laplace, 94010 Créteil Cedex, France*

* Corresponding author: ha.tran@lmd.jussieu.fr

**Abstract**

In this work, we measured the absorption by $CO_2$+$H_2O$ mixtures from 2400 to 2600 cm$^{-1}$ which corresponds to the spectral region beyond the $\nu_3$ band head of $CO_2$. Transmission spectra of $CO_2$ mixed with water vapor were recorded with a high-resolution Fourier-transform spectrometer for various pressure, temperature and concentration conditions. The continuum absorption by $CO_2$ due to the presence of water vapor was determined by subtracting from measured spectra the contribution of local lines of both species, that of the continuum of pure $CO_2$ as well as of the self- and $CO_2$-continua of water vapor induced by the $H_2O$-$H_2O$ and $H_2O$-$CO_2$ interactions. The obtained results are in very good agreement with the unique previous measurement (in a narrower spectral range). They confirm that the $H_2O$-continuum of $CO_2$ is significantly larger than that observed for pure $CO_2$. This continuum thus must be taken into account in radiative transfer calculations for media involving $CO_2$+$H_2O$ mixture. An empirical model, using sub-Lorentzian line shapes based on some temperature-dependent correction factors $\chi$ is proposed which enables an accurate description of the experimental results.

## 1. Introduction

Properly modeling the absorption spectrum of $CO_2$+$H_2O$ mixtures under various temperature and pressure conditions as well as for different gas concentrations is of great importance for planetary sciences. For instance, this is needed to explain the effect of $CO_2$ on the water vapor runaway greenhouse limit for Earth and other planets (Goldblatt et al., 2013; Popp et al., 2016; Ramirez et al., 2014; Turbet et al., 2016), a crucial point to understand why Venus and Earth had different fates. This also contributes to understand the future of Earth under the brightening Sun and more generally the habitability of extrasolar planets. For instance, water-rich extrasolar planets may lack the capability to regulate atmospheric $CO_2$, potentially leading to dense $CO_2$-$H_2O$ atmospheres (Wordsworth & Pierrehumbert 2013, Kitzmann et al. 2015, Marounina et al. 2017, Kite & Ford 2018). Following Ref. (Haberle et al., 2017), extreme events on early Mars could explain the geology of Mars (e.g. dry river beds and lakes) and mineralogy (e.g. clays). In particular, it has been proposed that meteoritic impact-generated steam atmosphere (made of large amounts of $CO_2$ and $H_2O$) could have induced episodic precipitations responsible for the formation of the Martian valley networks (Segura et al., 2012, 2008, 2002; Turbet et al., 2017). In this case, it is obvious that an accurate knowledge of the absorption spectrum of $CO_2+H_2O$ is essential. Such knowledge is also crucial to accurately model the evolution and observability of magma ocean planets, e.g. telluric planets that have surface temperatures high enough for their mantle to be in a liquid state, and that are expected to have outgassed large amounts of volatiles dominated by $H_2O$ and $CO_2$ (Abe and Matsui, 1988; Elkins-Tanton, 2008; Hamano et al., 2013; Lebrun et al., 2013; Lupu et al., 2014; Marcq, 2012; Marcq et al., 2017), assuming mantles relatively oxidizing as on present-day Earth and Venus. Modeling them properly serves to understand the early stage of the evolution of the Solar System rocky planets, as well as to anticipate and prepare future observations of young rocky extrasolar planets, or planets that recently suffered from a collision with a giant impactor.

Despite these potential applications for planetary-atmospheres studies, practically all studies devoted to spectra of $CO_2+H_2O$ mixtures are limited to spectroscopic parameters of isolated lines or local absorption. In fact, the infrared absorption spectrum of a $CO_2+H_2O$ mixture contains two different contributions. The first, called local absorption, is due to absorption in the center and near wings of the ro-vibrational lines of the monomer of each species. The second contribution is due to absorption by the stable and metastable dimers, to absorption induced by collisions and to absorption in the far wings of monomers lines. This contribution is often called "continuum absorption" in spectroscopy because of its smooth and slowly varying behavior with wavenumber (Hartmann et al., 2008). For local absorption by the monomers, half-width at half-maximum (HWHM) of several $H_2O$ lines broadened by $CO_2$ were measured and/or calculated in various studies [e.g. (Brown et al., 2007; Gamache et al., 2016; Lu et al., 2014; Poddar et al., 2009; Sagawa et al., 2009)] while $H_2O$-broadening coefficient of $CO_2$ lines were measured in (Delahaye et al., 2016; Sung et al., 2009). Local absorption by the monomers can then be computed using these broadening coefficients together with other spectroscopic line parameters such as the line positions and integrated intensities,… which are provided in various spectroscopic databases (Gordon et al., 2017; Jacquinet-Husson et al., 2016; Rothman et al., 2010). For continuum absorption, while several studies were devoted to the continua of pure $H_2O$ (or $CO_2$) as well as of $H_2O$ (or $CO_2$) in air [see (Baranov, 2011; Baranov et al., 2008; Clough et al., 1989; Hartmann, 1989; Hartmann et al., 2010, 1993; Hartmann and Perrin, 1989; Mlawer et al., 1999; Modelain et al., 2014; Perrin and Hartmann, 1989; Tran et al., 2011; Tretyakov et al., 2013; Mlawer et al., 2012), for instance], to the best of our knowledge, Ref. (Baranov, 2016) is the unique study dedicated to the measurement of the continuum absorption by $CO_2+H_2O$ mixtures. Using a Fourier-transform spectrometer and a multi-path cell, Y. I. Baranov (Baranov, 2016) measured transmission spectra of $CO_2+H_2O$ mixtures for various pressure, temperature and concentration conditions in the infrared. He established that at about 1100 $cm^{-1}$, the continuum absorption of $H_2O$ in $CO_2$ is nearly twenty times larger than that of $H_2O$ in $N_2$. This observation seems to be consistent with the theoretical results of Ma and Tipping (Ma and Tipping, 1992) where continuum absorption due to the far wings of $H_2O$ lines broadened by $CO_2$ and $N_2$ were calculated at room temperature between 0 and 10000 $cm^{-1}$. In Ref. (Baranov, 2016), it was also observed, for a limited spectral range in the far wing of the $CO_2$ $v_3$ band (from 2500 to about 2575 $cm^{-1}$) that the absorption of $CO_2$ in $H_2O$ is about one order of magnitude stronger than that of pure carbon dioxide. These results show that the $CO_2+H_2O$ continuum must be taken into account in the radiative transfer models for the various applications mentioned previously. Since continuum absorption strongly depends on the considered wavelength and absorption by $CO_2$ in $H_2O$ cannot be extrapolated from that of pure $CO_2$, a much larger spectral range for the $CO_2$ $\nu_3$ band wing is thus investigated in this work. The large spectral range considered also enables the development of an empirical model for the $H_2O$-continuum absorption of $CO_2$ in the $\nu_3$ band wing which could be easily used in applications.

In this paper, we first present an experimental study of the continuum absorption by $CO_2$ due to interaction with $H_2O$ in a region beyond the $CO_2$ $\nu_3$ band, from 2400 to 2600 cm$^{-1}$, much broader than that investigated in Ref. (Baranov, 2016). For this, we used a high-resolution Fourier-transform spectrometer and a White-type cell which can be heated to record about twenty $CO_2$+$H_2O$ spectra for various pressure, temperature and concentration conditions. The continuum absorption by $CO_2$ due to the presence of water vapor was then determined by subtracting from measured spectra the contribution of local lines of both species, that of the continuum of pure $CO_2$ as well as of the self- and $CO_2$-continua of $H_2O$. The obtained results are then compared with the previous measurements of Ref. (Baranov, 2016). In a second step, an empirical model is built in order to represent these experimentally determined values. It is based on a set of $\chi$-factors correcting the Lorentzian shape in the wings of the $H_2O$-broadened absorption lines of $CO_2$. This paper is organized as follows: the measurement procedure and data analysis are described in Sec. 2, the obtained results and the empirical model are presented and discussed in Sec. 3 while the main conclusions are drawn in Sec. 4.

## 2. Measurements procedure and data analysis

The high-resolution Fourier-transform spectrometer at LISA (Bruker IFS 120 HR) was used to record all spectra. The spectrometer was configured with a globar as the broad-band light source, a KBr beam splitter and an InSb detector. The unapodized spectral resolution of 0.1 cm$^{-1}$, corresponding to a maximum optical path difference of 9 cm, was used for all measured spectra. The diameter of the FTS iris aperture was set to 2 mm. A White-type absorption cell, made of Pyrex glass and equipped with wedged $CaF_2$ windows was connected to the FTS with a dedicated optical interface inside the sample chamber of the FTS. Its base length is 0.20 m and, for the experiments described here, an optical path of 7.20 m was used. This cell can be heated to temperatures up to 100°C with a variation of 0.5°C along the cell, as measured with a type-K thermocouple (±1.5°C). In order to avoid condensation and to be able to work with significant $H_2O$ pressures, the cell and the entire gas-handling system (including the pressure gauges) were enclosed inside a thermally insulated Plexiglas box. The temperature inside the box is regulated by an air heating system at a temperature of about 60°C. The gas pressure was measured using three capacitive pressure transducers with 100 and 1000 Torr (1 Torr = 1.333 mbar) full scales, with a stated accuracy of ±0.12%. The spectral coverage from 1000 to 4500 cm$^{-1}$ was recorded for all measurements. The experiments were carried out as follows: Firstly, the temperature in the cell and that in the box were set to the desired values. Then when these temperatures were stabilized (after about 1 hour for the box and 5 hours for the cell), a spectrum was first recorded with the empty cell to provide the 100% transmission. The cell was then filled with about 760 Torr of $CO_2$ and a pure $CO_2$ spectrum was recorded. After being pumped out again, the cell was filled with water vapor, purified by several distillations, at the desired pressure (varying from 40 to 110 Torr). Then, $CO_2$ was introduced until the total pressure reaches a given value (from 380 to 760 Torr). Once the sample was well mixed, a spectrum was recorded using an averaging of 200 scans providing a signal-to-noise ratio of about 500 (RMS) for a recording duration of 16 minutes. The temperature and pressure in the cell were simultaneously recorded every 5 s. This showed that the temperature and pressure variations during the recording of a spectrum remained lower than 0.2 K and 0.5 Torr, respectively. The pressure and temperature conditions for all measurements are summarized in Table 1. Transmission spectra were obtained by dividing the spectra recorded with the gas sample by that obtained with the empty cell.

| Spectrum | Temperature (K) | H$_2$O pressure (Torr) | CO$_2$ pressure (Torr) |
|---|---|---|---|
| 1 | 367.15 | 0 | 760.6 |
| 2 | 366.65 | 108.78 | 760.25 |
| 3 | 366.45 | 86.78 | 606.6 |
| 4 | 366.55 | 54.54 | 381.3 |
| 5 | 366.35 | 109.90 | 608.9 |
| 6 | 366.35 | 68.95 | 380.9 |
| 7 | 364.65 | 108.90 | 381.45 |
| 8 | 344.95 | 0 | 761.55 |
| 9 | 344.65 | 103.90 | 764.55 |
| 10 | 344.64 | 82.65 | 608.15 |
| 11 | 344.65 | 52.15 | 382.95 |
| 12 | 344.58 | 105.60 | 607.90 |
| 13 | 344.60 | 66.19 | 380.55 |
| 14 | 344.50 | 105.10 | 381.55 |
| 15 | 325.15 | 0 | 759.4 |
| 16 | 325.18 | 79.00 | 761.8 |
| 17 | 325.19 | 63.19 | 609.1 |
| 18 | 325.15 | 39.47 | 380.6 |
| 19 | 325.35 | 80.20 | 610.3 |
| 20 | 325.25 | 50.17 | 382.4 |
| 21 | 325.15 | 80.30 | 382.9 |

Table 1: Experimental conditions of the measured spectra. The path length (*L*) used for all measurements was fixed to 7.20 m.

The total absorption coefficient (i.e. $\alpha$ in cm$^{-1}$) at wavenumber $\sigma$ (cm$^{-1}$) of a CO$_2$-H$_2$O mixture of temperature *T* (in Kelvin), total density $\rho_{tot}$ (in amagat) and mole fractions $x_{CO_2}$ and $x_{H_2O}$ can be written as:

$$\alpha(\sigma, x_{CO_2}, x_{H_2O}, \rho_{tot}, T) = \sum_{X=CO_2, H_2O} \alpha_{local}^X(\sigma, \Delta\sigma_X, x_{CO_2}, x_{H_2O}, \rho_{tot}, T)$$

$$+ \sum_{X=CO_2, H_2O} \sum_{Y=CO_2, H_2O} \alpha_{CA}^{X-Y}(\sigma, \Delta\sigma_X, x_{CO_2}, x_{H_2O}, \rho_{tot}, T)$$

(1)

where $\alpha_{local}^X(\sigma, \Delta\sigma_X, x_{CO_2}, x_{H_2O}, \rho_{tot}, T)$ denotes the absorption due to local lines of the monomer X whose extensions are limited to $\pm\Delta\sigma_X$ around the line center and $\alpha_{CA}^{X-Y}(\sigma, \Delta\sigma_X, x_{CO_2}, x_{H_2O}, \rho_{tot}, T)$ is the continuum absorption due to species X interacting with species Y. Provided that $\Delta\sigma_X$ is much greater than the widths of the lines of species X under the considered T and P conditions, one can write (Hartmann et al., 2008):

$$\alpha_{CA}^{X-Y}(\sigma, \Delta\sigma_X, x_{CO_2}, x_{H_2O}, \rho_{tot}, T) = \rho_{tot}^2 x_{CO_2} x_{H_2O} \, CA_{X-Y}(\sigma, \Delta\sigma_X, T), \qquad (2)$$

where $CA_{X-Y}$ (in cm$^{-1}$/amagat$^2$) is the squared-density normalized continuum absorption due to molecule X "influenced" by the presence of molecule Y. The possible origin of the continua will be discussed in the next section.

In order to deduce $CA_{CO_2-H_2O}$ from the measured spectra, the following procedure was used: (i) $\alpha_{local}^{CO_2}$ and $\alpha_{local}^{H_2O}$ were calculated by using spectroscopic data given in the 2012 version of the HITRAN database (Rothman et al., 2013) for the line positions and integrated intensities, the energies of the lower levels of the transitions and the self-broadening coefficients (i.e. the pressure-normalized HWHMs). The $H_2O$-broadening coefficients of $CO_2$ lines as well as their temperature dependences were calculated following the analytical formulation proposed in Ref. (Sung et al., 2009). The $CO_2$-broadening coefficients of $H_2O$ lines were scaled from those of air, as done in Ref. (Baranov, 2016), their temperature dependences being fixed to those of air (Rothman et al., 2013). In the absence of available data, the needed $CO_2$ and $H_2O$ pressure shifts were assumed to be the same as the air-induced ones, provided by the HITRAN database (Rothman et al., 2013). The temperature dependences of the self-broadening coefficients for $CO_2$ and $H_2O$ lines were also set to be the same as those of the air-broadening coefficients. Since the relative contribution of the local lines is quite small, these approximations lead to very small changes of the total absorptions and do not affect the deduced values of $CA_{CO_2-H_2O}$. The influence of the apparatus line-shape function was also taken into account by convolving the calculated transmission (i.e. $T^{calc}(\sigma, x_{CO_2}, x_{H_2O}, \rho_{tot}, T) = \exp[-L\alpha(\sigma, x_{CO_2}, x_{H_2O}, \rho_{tot}, T)]$) with an instrument line shape accounting for the finite maximum optical path difference as well as the iris radius. The contribution of each $H_2O$ line to $\alpha_{local}^{H_2O}$ was calculated between -25 and 25 cm$^{-1}$ away from the line center (i.e. $\Delta\sigma_{H_2O}$ = 25 cm$^{-1}$), in order to be consistent with the choice adopted for the water vapor continua $CA_{H_2O-H_2O}$ (Clough et al., 1989; Mlawer et al., 2012) and $CA_{H_2O-CO_2}$ (Ma and Tipping, 1992; Pollack et al., 1993). For $CO_2$ lines, $\Delta\sigma_{CO_2}$ = 5 cm$^{-1}$ was used in the computation of $\alpha_{local}^{CO_2}$.

The contributions of the continua of pure $H_2O$ and $H_2O$ in $CO_2$ were calculated as follows. Absorption by the self-continuum ($CA_{H_2O-H_2O}$) of $H_2O$ was taken from the MT_CKD 3.0 database (Mlawer et al., 2012), available on http://rtweb.aer.com/. The $CO_2$- continuum of $H_2O$ ($CA_{H_2O-CO_2}$) was calculated with the line shape correction functions $\chi$ of Ref. (Ma and Tipping, 1992) using line positions and intensities from the 2012 version of the HITRAN database (Rothman et al., 2013) with a cut-off at 25 cm$^{-1}$ to remove the local line contribution. Its temperature dependence was empirically derived using data provided in Ref. (Pollack et al., 1993).

The absorption due to the self-continuum of $CO_2$, i.e. $CA_{CO_2-CO_2}$, was taken from Ref. (Tran et al., 2011) in which absorption of pure $CO_2$ beyond the $\nu_3$ band head was measured at temperatures from 260 to 473 K. The values of $CA_{CO_2-CO_2}$ under the temperature conditions considered in the present study were then deduced from those of Tran et al (Tran et al., 2011) using a linear interpolation in temperature. The obtained values were compared with those directly deduced from the present measurements (i.e. Spectra number 1, 8 and 15 in Table 1) for pure $CO_2$ showing very good agreements.

## 3. Results

Figure 1 presents an example of the absorption coefficient (black) of a $CO_2$-$H_2O$ mixture measured at 325.18 K and for a total pressure of 761.8 Torr, the molar fraction of $H_2O$ in the mixture being 0.1037 (spectrum 16 as referred in Table 1). The calculated contributions of local $H_2O$ and $CO_2$ lines, ($\alpha_{local}^{CO_2} + \alpha_{local}^{H_2O}$) (red line), of the self- (green) and $CO_2$- (blue) continua of $H_2O$ (i.e. $\alpha_{CA}^{H_2O-H_2O}$ and $\alpha_{CA}^{H_2O-CO_2}$) and that of absorption due to the self-continuum of $CO_2$ (i.e. $\alpha_{CA}^{CO_2-CO_2}$, cyan) are also plotted on this figure. Following Eqs. (1,2), the difference between the measured absorption coefficient and the sum of all these contributions directly yields the absorption due to the continuum of $CO_2$ in $H_2O$ (i.e. $\alpha_{CA}^{CO_2-H_2O}$, olive). As can be seen on this figure, the relative contribution of the self- and $CO_2$-continua of $H_2O$ to the total absorption is small and absorption is mainly due to the self- and $H_2O$-continua of $CO_2$. Therefore, uncertainties of the self- and $CO_2$-continua of $H_2O$ will not significantly affect the obtained result. The local lines contribution is correctly reproduced by the calculation leading to a smooth behavior of the values of $\alpha_{CA}^{CO_2-H_2O}$ obtained from the above-described procedure. This treatment was applied to all measured spectra, yielding a set of values of $\alpha_{CA}^{CO_2-H_2O}$ for various mixtures and pressure and temperature conditions of the recorded spectra (see Table 1).

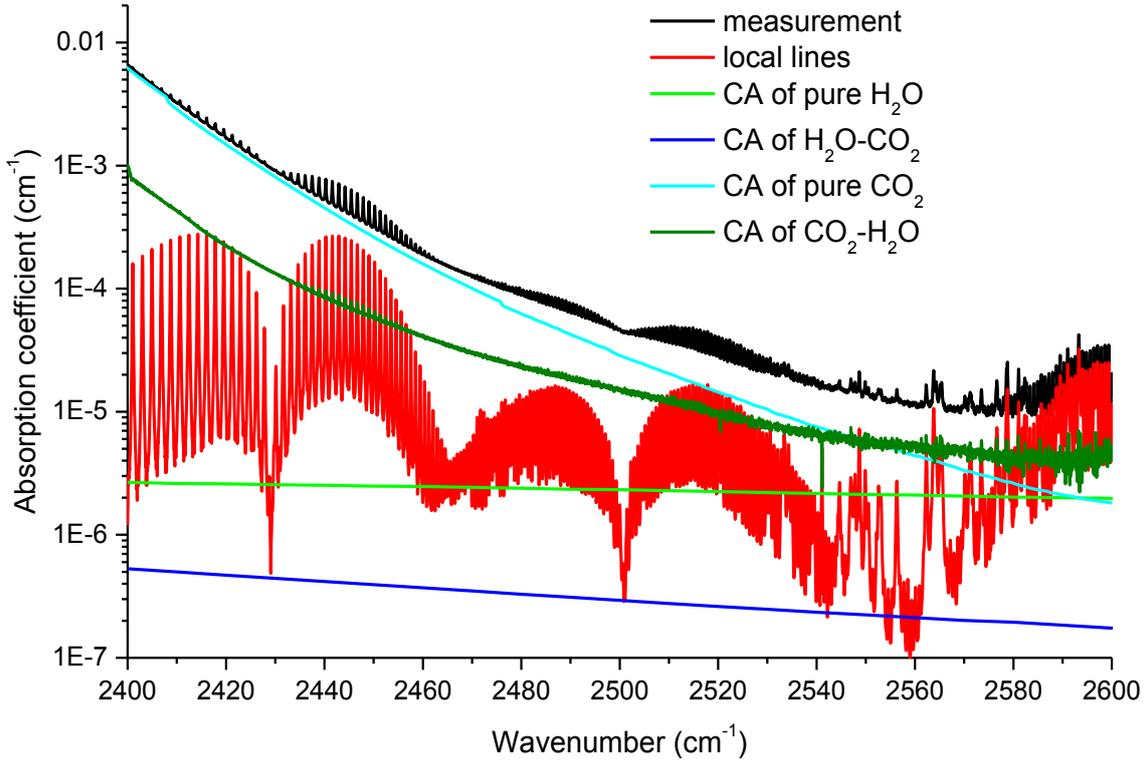

Figure 1: Example of the absorption coefficient of a $CO_2$-$H_2O$ mixture measured at 325.18 K and 761.8 Torr with a molar fraction of 0.1037 for $H_2O$. In red is the calculated contributions of local lines of $CO_2$ and $H_2O$ while in green and blue are those due to the self- (Mlawer et al., 2012) and $CO_2$-continua (Ma and Tipping, 1992; Pollack et al., 1993) of $H_2O$, respectively. Absorption due to the self-continuum of $CO_2$ is represented by the cyan curve. All these contributions are subtracted from the measurement to deduce the contribution of continuum absorption of $CO_2$ broadened by $H_2O$ (olive).

Figure 2 shows examples of the dependence of $\alpha_{CA}^{CO_2-H_2O}$ on the product of the $H_2O$ and $CO_2$ densities, i.e. $\rho_{tot}^2\, x_{CO_2} x_{H_2O}$ for two wavenumbers 2461.57 and 2508.99 cm$^{-1}$. As can be observed, nice linear dependences are obtained, in agreement with Eq. (2). The slope of a linear fit thus directly yields $CA_{CO_2-H_2O}$ [see Eq. (2)], leading to 6.32 x 10$^{-4}$ (±0.05 x 10$^{-4}$) and 2.03 x 10$^{-4}$ (±0.08 x 10$^{-4}$) cm$^{-1}$/amagat$^2$ for σ = 2461.57 and 2508.99 cm$^{-1}$, respectively.

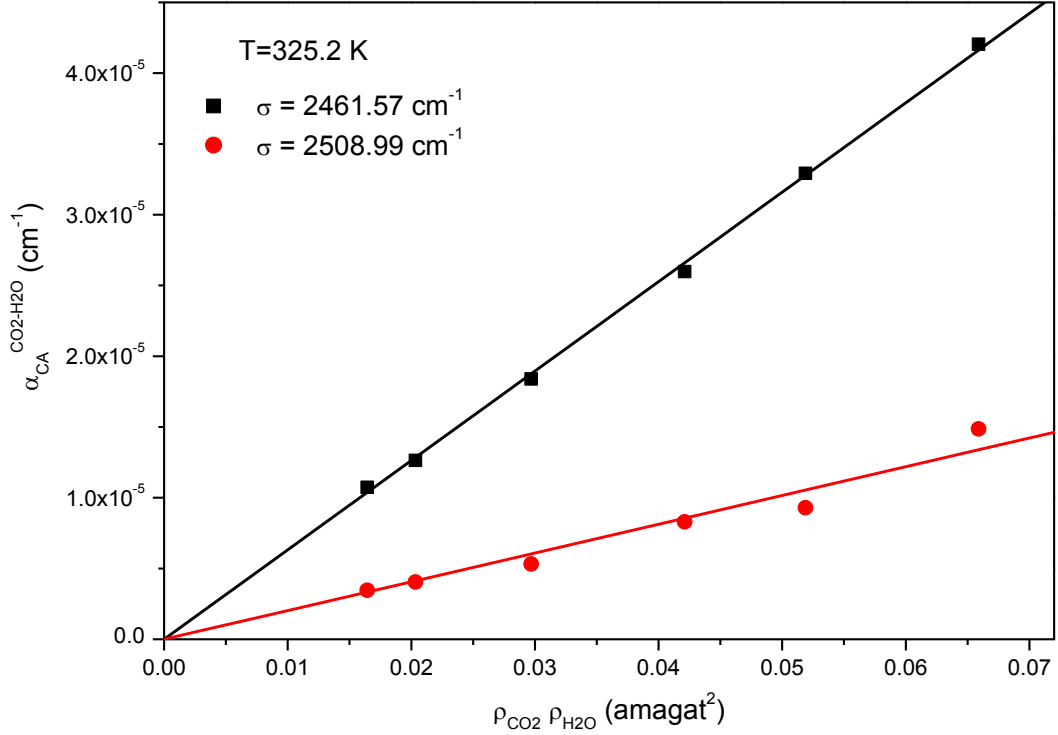

Figure 2: Dependences of $\alpha_{CA}^{CO_2-H_2O}$ on the product of the H$_2$O and CO$_2$ densities (i.e. $\rho_{CO_2}\rho_{H_2O} = \rho_{tot}^2\, x_{CO_2} x_{H_2O}$) for two wavenumbers, deduced from measurements at 325.2 K and their linear fits.

Experimental values of $CA_{CO_2-H_2O}$, deduced as explained above in all the investigated spectral region are plotted in Figure 3 (black points). These values were averaged over all measured temperatures since no clear temperature dependence could be observed within the studied temperature range, as it was the case in Ref. (Baranov, 2016). This indicates that the temperature dependence, if any, must be small as it was shown to be the case, for a 50 K broad temperature interval, for the self- (Hartmann and Perrin, 1989), N$_2$ (Perrin and Hartmann, 1989) and Ar- (Boissoles et al., 1989) continua of CO$_2$ in the same region. The plotted uncertainties (Fig. 3) correspond to the standard deviation of the linear fits (Fig. 2) and of the temperature average. For comparison, the values measured in Ref. (Baranov, 2016) were also plotted (red points) in this figure, showing a very good agreement. The values of $CA_{CO_2-H_2O}$ are listed in the supplementary material file.

The origin of the continuum absorption by CO$_2$ in the region beyond the ν$_3$ band head is not fully clear. Indeed, the contributions of the far wing of the lines due to the intrinsic (vibrating) dipole of the CO$_2$ molecules, of the collision-induced dipole and of stable and meta-stable dimers all show a linear dependence versus the squared total density (or ρ$_X$ρ$_Y$ product), as the observed one (see Fig. 2). In fact, while it was though for a long time (Boissoles et al., 1989; Hartmann and Perrin, 1989; Perrin and Hartmann, 1989; Tipping et al., 1999) that only the first mechanism was involved, it was recently shown that the transient dipoles induced in interacting molecular pair, plays a role (Hartmann and Boulet, 2011). Solving this issue in the case of $CO_2$-$H_2O$ is a vast and complex problem that is currently under study. However, there is a need for computational tools suitable for applications such as the ones mentioned in the introduction of this paper. Within this frame, and although this may not be fully rigorous from the point of view of physics, the widely-used $\chi$-factor approach [see Refs. (Perrin and Hartmann, 1989; Tran et al., 2011; Turbet and Tran, 2017) for instance] seems to be a good compromise. It connects the observed absorption to contributions of the lines due to the intrinsic dipole of the monomer and allows to accurately represent the observations as shown in the above-mentioned references and by the results below. Besides, it can be used to model the contribution of local lines and for extrapolations to other spectral regions, which may be risky but is the only solution in many cases due to the absence of any other model or data. Within this approach, the absorption from the centers to the far wings of the lines of species X in a mixture with species Y, is calculated using the following equation:

$$\alpha^{X-Y}(\sigma, x_X, x_Y, \rho_{tot}, T)$$
$$= \rho_{tot} x_X \sum_i S_i(T) \exp\left[\frac{hc(\sigma - \sigma_i)}{2k_B T}\right] \times \frac{1 - \exp(-\frac{hc\sigma}{k_B T})}{1 - \exp(-\frac{hc\sigma_i}{k_B T})} \times \frac{\sigma}{\sigma_i} \times \frac{1}{\pi}$$
$$\times \sum_{Pert=X,Y} \frac{\Gamma_i^{X-Pert}(T) \chi^{X-Pert}(T, |\sigma - \sigma_i|)}{[\sigma - \sigma_i - \Delta_i^{X-Pert}(T)]^2 + [\Gamma_i^{X-Pert}(T)]^2}$$

(3)

where $x_X$ and $x_Y$ are the molar fractions of species X and Y, respectively. The sums extend over all the lines of species X contributing to the absorption at the current wavenumber σ. The $\exp\left[\frac{hc(\sigma-\sigma_i)}{2k_B T}\right]$ term is the quantum asymmetry factor resulting from the so-called fluctuation-dissipation theorem (Hartmann et al., 2008). $\sigma_i, S_i(T), \Gamma_i^{X-Pert}$ and $\Delta_i^{X-Pert}$ are respectively the unperturbed line position (cm$^{-1}$), integrated line intensity (cm$^{-2}$.amagat$^{-1}$), the line width and shift (both in cm$^{-1}$) due to collisions of the active molecule $X$ with the perturbator $Pert$. The $\sigma[1 - \exp\left(-\frac{hc\sigma}{k_B T}\right)]$ term is related to spontaneous emission at wavenumber $\sigma$. The line-shape correction factor $\chi^{X-Pert}(T, |\sigma - \sigma_i|)$ is assumed to be independent of the transition.

From this general equation, the continuum absorption of $CO_2$ in $H_2O$ (i.e. absorption in the far wings of $CO_2$ lines broadened by $H_2O$ within this approach) can be expressed as:

$$CA_{CO_2-H_2O}(\sigma, T)$$
$$= \sum_i S_i(T) \exp\left[\frac{hc(\sigma - \sigma_i)}{2k_B T}\right] \times \frac{1 - \exp(-\frac{hc\sigma}{k_B T})}{1 - \exp(-\frac{hc\sigma_i}{k_B T})} \times \frac{\sigma}{\sigma_i} \times \frac{1}{\pi}$$
$$\times \frac{\gamma_i^{CO_2-H_2O}(T) \chi^{CO_2-H_2O}(T, |\sigma - \sigma_i|)}{[\sigma - \sigma_i]^2}$$

(4)

where the sum is now restricted to the lines centered outside the [$(\sigma - 5)$ and $(\sigma + 5)$] cm$^{-1}$ range and $\gamma_i^{CO_2-H_2O}(T)$ is the $H_2O$-broadening coefficient (cm$^{-1}$/amagat) of $CO_2$ lines. The temperature-dependent $\chi^{CO2-H2O}$ factors were thus determined by fitting this equation to the measured values of $CA_{CO_2-H_2O}$ (Fig. 3). A functional form for the $\chi^{CO2-H2O}$ factors, similar to what was constructed for pure $CO_2$ in Refs. (Hartmann and Perrin, 1989; Perrin and Hartmann, 1989; Tran et al., 2011) was adopted in this work, i.e.:

$$
\begin{aligned}
0 < \Delta\sigma \leq \sigma_1 \quad & \chi(T,\Delta\sigma) = 1 \\
\sigma_1 < \Delta\sigma \leq \sigma_2 \quad & \chi(T,\Delta\sigma) = \exp[-B_1(\Delta\sigma - \sigma_1)] \\
\sigma_2 < \Delta\sigma \leq \sigma_3 \quad & \chi(T,\Delta\sigma) = \exp[-B_1(\sigma_2 - \sigma_1) - B_2(\Delta\sigma - \sigma_2)] \\
\sigma_3 < \Delta\sigma \quad & \chi(T,\Delta\sigma) = \exp[-B_1(\sigma_2 - \sigma_1) - B_2(\sigma_3 - \sigma_2) - B_3(\Delta\sigma - \sigma_3)]
\end{aligned}
\tag{5}
$$

The temperature dependences of the parameters $B_i$ were determined such that the density-squared normalized absorption coefficients [i.e. $CA_{CO_2-H_2O}$ in Eq. (1)] in the entire region 2400-2600 cm$^{-1}$ are temperature-independent. This was done for the 200-500 K temperature range. The values of $\sigma_1, \sigma_2$ and $\sigma_3$ as well as $B_1, B_2$ and $B_3$ were determined by fitting Eqs. (4) and (5) on the measured values of $CA_{CO_2-H_2O}$ (Fig. 3), i.e.:

$$
\sigma_1 = 5;\ \sigma_2 = 35\ \text{and}\ \sigma_3 = 170\ \text{cm}^{-1},
$$

$$
B_1 = 0.0689 - \frac{2.4486}{T} + 64.085/T^2, \tag{6}
$$

$$
B_2 = 0.00624 + \frac{3.7273}{T} - 299.144/T^2,
$$

$$
\text{and}\ B_3 = 0.0025\ .
$$

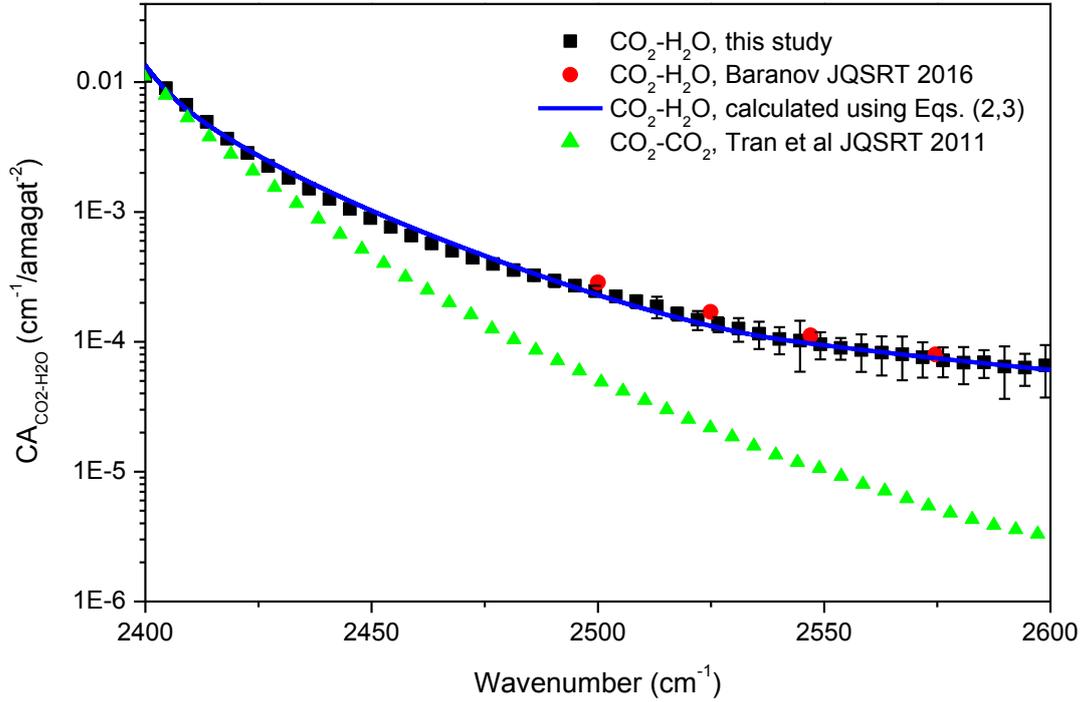

Figure 3: Continuum absorption of $CO_2$ broadened by $H_2O$, $CA_{CO_2-H_2O}$, beyond the $CO_2$ $\nu_3$ band head region measured in this work (black rectangles) and those measured by Ref. (Baranov, 2016) (red circles). Values of $CA_{CO_2-H_2O}$ calculated from the sub-Lorentzian empirical model [Eqs. (4-6)] are represented by the blue line. The self-continuum of $CO_2$ (Tran et al., 2011) are also plotted (green) for comparison.

The quality of the fit is demonstrated in Fig. 3 where the absorption coefficients calculated using Eqs. (4-6) (blue line) are in very good agreement with the experimental values (black points). These temperature-dependent $\chi$-factors [Eqs. (5,6)] can now be used to model $H_2O$-broadened $CO_2$ far line wings in applications such as those mentioned in Sec. 1.

In Ref. (Baranov, 2016), it was shown that $CA_{CO_2-H_2O}$ is about one order of magnitude stronger than that in pure $CO_2$ between 2500 and 2575 cm$^{-1}$. Since the present study covers a significantly broader spectral range, we verify this by comparing $CA_{CO_2-H_2O}$ with $CA_{CO_2-CO_2}$ for the whole considered spectral region. For that, we plot in Fig. 3 the values of $CA_{CO_2-CO_2}$, measured at room temperature by Tran et al (Tran et al., 2011) (green points). This figure confirms that the values of $CA_{CO_2-H_2O}$ are indeed significantly larger than those of $CA_{CO_2-CO_2}$, but their ratio is not constant and increases with the wavenumber. This may be qualitatively explained by the effect of incomplete collisions. In fact, in Ref. (Tran et al., 2017) it was shown that incomplete collisions (i.e. collisions that are ongoing or start at time zero) lead to an increase of absorption in the line wings. Since the $CO_2$-$H_2O$ intermolecular potential involves much larger long-range contributions than that of $CO_2$-$CO_2$, the effect of incomplete collisions must be stronger for $CO_2$ in $H_2O$ than for pure $CO_2$. This explanation is also consistent with the observed relative magnitudes of the continua of $CO_2$ in $N_2$ (Perrin and Hartmann, 1989), Ar (Boissoles et al., 1989) and He (Ozanne et al., 1995) [see also Fig. 6 of Ref. (Baranov, 2016)]. However, since line-mixing effects (Tran et al., 2011) [but likely also the collision-induced dipole moment (Hartmann and Boulet, 2011)] contribute to absorption in this spectral region, explaining its behavior as well as analyzing its origin are beyond the scope of this paper and will be carried out in a future study.

## 4. Conclusion

Absorption in the spectral region beyond the 4.3 μm ($\nu_3$) band of $CO_2$ broadened by $H_2O$ was measured with a high-resolution Fourier-transform spectrometer under various pressure and temperature conditions. The measured values are in very good agreement with the unique previous measurement but extend the investigated spectral range. The results show that the $CO_2$+$H_2O$ absorption continuum in this spectral region is significantly larger than the pure $CO_2$ continuum. Therefore, this continuum must be taken into account in radiative transfer calculations for media involving $CO_2$+$H_2O$ mixture. An empirical model, using sub-Lorentzian line shapes based on temperature-dependent $\chi$-factors was then deduced from the measured values, enabling easy calculations of absorption in the $\nu_3$ band wing of $CO_2$ broadened by $H_2O$. The measurements presented in our manuscript are part of a broader project aiming at characterizing several absorption properties of $CO_2$+$H_2O$ mixtures (Turbet et al., 2017). The effect of these new measurements on various planetary environments will be quantitatively investigated in a future, dedicated study.

**Acknowledgment**

*The authors thank Dr. Q. Ma for providing his calculated data of the $CO_2$-continuum of water vapor at various temperatures. J.-M. Hartmann is acknowledged for helpful discussions.*

**References**

Abe, Y., Matsui, T., 1988. Atmosphere and Formation of a Hot Proto-Ocean on Earth. J. Atmos. Sci. doi:10.1175/1520-0469(1988)045<3081:EOAIGH>2.0.CO;2

Baranov, Y.I., 2016. On the significant enhancement of the continuum-collision induced absorption in $H_2O$+$CO_2$ mixtures. J. Quant. Spectrosc. Radiat. Transf. 175, 100–106. doi:10.1016/j.jqsrt.2016.02.017

Baranov, Y.I., 2011. The continuum absorption in $H_2O$+$N_2$ mixutres in the 2000-3250 $cm^{-1}$ spectral region at temperatures from 326 to 363 K. J. Quant. Spectrosc. Radiat. Transf. 112, 2281–2286. doi:10.1016/j.jqsrt.2011.01.024

Baranov, Y.I., Lafferty, W.J., Ma, Q., Tipping, R.H., 2008. Water-vapor continuum absorption in the 800-1250 $cm^{-1}$ spectral region at temperatures from 311 to 363 K. J. Quant. Spectrosc. Radiat. Transf. 109, 2291–2302. doi:10.1016/j.jqsrt.2008.03.004

Boissoles, J., Menoux, V., Le Doucen, R., Boulet, C., Robert, D., 1989. Collisionally induced population transfer effect in infrared absorption spectra. II. The wing of the Ar-

broadened $\nu_3$ band of $CO_2$. J. Chem. Phys. 91, 2163. doi:10.1063/1.457024

Brown, L.R., Humphrey, C.M., Gamache, R.R., 2007. $CO_2$-broadened water in the pure rotation and $\nu_2$ fundamental regions. J. Mol. Spectrosc. 246, 1–21. doi:10.1016/j.jms.2007.07.010

Clough, S.A., Kneizys, F.X., Davies, R.W., 1989. Line shape and the water vapor continuum. Atmos. Res. 23, 229–241. doi:10.1016/0169-8095(89)90020-3

Delahaye, T., Landsheere, X., Pangui, E., Huet, F., Hartmann, J.-M., Tran, H., 2016. Broadening of $CO_2$ lines in the 4.3μm region by $H_2O$. J. Mol. Spectrosc. 326, 17–20. doi:10.1016/j.jms.2016.02.007

Elkins-Tanton, L.T., 2008. Linked magma ocean solidification and atmospheric growth for Earth and Mars. Earth Planet. Sci. Lett. 271, 181–191. doi:10.1016/j.epsl.2008.03.062

Gamache, R.R., Farese, M., Renaud, C.L., 2016. A spectral line list for water isotopologues in the 1100–4100 cm$^{-1}$ region for application to $CO_2$-rich planetary atmospheres. J. Mol. Spectrosc. 326, 144–150. doi:10.1016/j.jms.2015.09.001

Goldblatt, C., Robinson, T.D., Zahnle, K.J., Crisp, D., 2013. Low simulated radiation limit for runaway greenhouse climates. Nat. Geosci. 6, 661–667. doi:10.1038/ngeo1892

Gordon, I., Rothman, L., Hill, C., Kochanov, R., Tan, Y., Bernath, P., Boudon, V., Campargue, A., Drouin, B., Flaud, J.M., Gamache, R., Hodges, J., Perevalov, V., Shine, K., Smith, M., 2017. The HITRAN2016 Molecular Spectroscopic Database. J Quant Spectrosc Radiat Transf, in press. doi:10.1016/j.jqsrt.2017.06.038

Haberle, R., Catling, D., Carr, M., & Zahnle, K., 2017. The Early Mars Climate System, in: R. Haberle, R. Clancy, F. Forget, M. Smith, & R.Z. (Ed.), The Atmosphere and Climate of Mars. Cambridge University Press, Cambridge, pp. 497–525.

Hamano, K., Abe, Y., Genda, H., 2013. Emergence of two types of terrestrial planet on solidification of magma ocean. Nature 497, 607–610. doi:10.1038/nature12163

Hartmann, J.-M., 1989. Measurements and calculations of $CO_2$ room-temperature high-pressure spectra in the 4.3 μm region. J. Chem. Phys. 90, 2944–2950. doi:10.1017/CBO9781107415324.004

Hartmann, J.-M., Boulet, C., Robert, D., 2008. Collisional effects on molecular spectra. Laboratory experiments and models, consequences for applications. Elsevier, Amsterdam.

Hartmann, J.-M., Boulet, C., Tran, H., Nguyen, M.T., 2010. Molecular dynamics simulations for $CO_2$ absorption spectra. I. Line broadening and the far wing of the $\nu_3$ infrared band. J. Chem. Phys. 133, 144313. doi:10.1063/1.3489349

Hartmann, J.M., Boulet, C., 2011. Molecular dynamics simulations for $CO_2$ spectra. III. Permanent and collision-induced tensors contributions to light absorption and scattering. J. Chem. Phys. 134, 184312. doi:10.1063/1.3589143

Hartmann, J.M., Perrin, M.Y., 1989. Measurements of pure $CO_2$ absorption beyond the $\nu_3$ bandhead at high temperature. Appl. Opt. 28, 2550–3. doi:10.1364/AO.28.002550

Hartmann, J.M., Perrin, M.Y., Ma, Q., Tippings, R.H., 1993. The infrared continuum of pure water vapor: Calculations and high-temperature measurements. J. Quant. Spectrosc.

Radiat. Transf. 49, 675–691. doi:10.1016/0022-4073(93)90010-F

Jacquinet-Husson, N., Armante, R., Scott, N.A., Chédin, A., Crépeau, L., Boutammine, C., Bouhdaoui, A., Crevoisier, C., Capelle, V., Boonne, C., Poulet-Crovisier, N., Barbe, A., Chris Benner, D., Boudon, V., Brown, L.R., Buldyreva, J., Campargue, A., Coudert, L.H., Devi, V.M., Down, M.J., Drouin, B.J., Fayt, A., Fittschen, C., Flaud, J.M., Gamache, R.R., Harrison, J.J., Hill, C., Hodnebrog, Hu, S.M., Jacquemart, D., Jolly, A., Jiménez, E., Lavrentieva, N.N., Liu, A.W., Lodi, L., Lyulin, O.M., Massie, S.T., Mikhailenko, S., Müller, H.S.P., Naumenko, O. V., Nikitin, A., Nielsen, C.J., Orphal, J., Perevalov, V.I., Perrin, A., Polovtseva, E., Predoi-Cross, A., Rotger, M., Ruth, A.A., Yu, S.S., Sung, K., Tashkun, S.A., Tennyson, J., Tyuterev, V.G., Vander Auwera, J., Voronin, B.A., Makie, A., 2016. The 2015 edition of the GEISA spectroscopic database. J. Mol. Spectrosc. 327, 31–72. doi:10.1016/j.jms.2016.06.007

Kite, Edwin S.; Ford, Eric B., 2018. Habitability of exoplanet waterworlds, eprint arXiv:1801.00748.

Kitzmann, D.; Alibert, Y.; Godolt, M.; Grenfell, J. L.; Heng, K.; Patzer, A. B. C.; Rauer, H.; Stracke, B.; von Paris, P., 2015. The unstable CO2 feedback cycle on ocean planets. Monthly Notices of the Royal Astronomical Society, Volume 452, Issue 4, p.3752-3758. https://doi.org/10.1093/mnras/stv1487

Lebrun, T., Massol, H., Chassefière, E., Davaille, A., Marcq, E., Sarda, P., Leblanc, F., Brandeis, G., 2013. Thermal evolution of an early magma ocean in interaction with the atmosphere. J. Geophys. Res. E Planets 118, 1155–1176. doi:10.1002/jgre.20068

Lu, Y., Li, X.F., Liu, A.W., Hu, S.M., 2014. $CO_2$ pressure shift and broadening of water lines near 790 nm. Chinese J. Chem. Phys. 27, 1–4. doi:10.1063/1674-0068/27/01/1-4

Lupu, R.E., Zahnle, K., Marley, M.S., Schaefer, L., Fegley, B., Morley, C., Cahoy, K., Freedman, R., Fortney, J.J., 2014. the Atmospheres of Earthlike Planets After Giant Impact Events. Astrophys. J. 784, 27. doi:10.1088/0004-637X/784/1/27

Ma, Q., Tipping, R.H., 1992. A far wing line shape theory and its application to the foreign-broadened water continuum absorption. III. J. Chem. Phys. 97, 818–828. doi:10.1063/1.463184

Marcq, E., 2012. A simple 1-D radiative-convective atmospheric model designed for integration into coupled models of magma ocean planets. J. Geophys. Res. E Planets 117, 1–10. doi:10.1029/2011JE003912

Marcq, E., Salvador, A., Massol, H., Davaille, A., 2017. Thermal radiation of magma ocean planets using a 1-D radiative-convective model of $H_2O$-$CO_2$ atmospheres. J. Geophys. Res. Planets 122, 1539–1553. doi:10.1002/2016JE005224

Marounina, N.; Rogers, L. A.; Kempton, E., 2017. Constraining the Habitable Zone Boundaries for Water World Exoplanets. Habitable Worlds 2017: A System Science Workshop, held 13-17 November, 2017 in Laramie, Wyoming. LPI Contribution No. 2042, id.4135.

Mlawer, E.J., Clough, S.A., Brown, P.D., Tobin, D., 1999. Recent Developments in the Water Vapor Continuum Observations and the CKD Continuum Model. Ninth ARM Sci. Team Meet. Proc. 2, 1–6.

Mlawer, E.J., Payne, V.H., Moncet, J.-L., Delamere, J.S., Alvarado, M.J., Tobin, D.C., 2012.

Development and recent evaluation of the MT_CKD model of continuum absorption. Philos. Trans. R. Soc. A Math. Phys. Eng. Sci. 370, 2520–2556. doi:10.1098/rsta.2011.0295

Modelain, D., Manigand, S., Kassi, S., Campargue, A., 2014. Temperature dependence of the water vapor self-continuum by cavity ring-down spectroscopy in the 1.6μm transparency window. J. Geophys. Res. Atmos. 5625–5639. doi:10.1002/2013JD021319.Received

Ozanne, L., Nguyen, V.T., Brodbeck, C., Bouanich, J.-P., Hartmann, J.-M., Boulet, C., 1995. Line mixing and nonlinear density effects in the $\nu_3$ and $3\nu_3$ infrared bands of $CO_2$ perturbed by He up to 1000 bar. J. Chem. Phys 102, 7306–7316.

Perrin, M.Y., Hartmann, J.M., 1989. Temperature-dependent measurements and modeling of absorption by $CO_2$-$N_2$ mixtures in the far line-wings of the 4.3μm $CO_2$ band. J. Quant. Spectrosc. Radiat. Transf. 42, 311–317. doi:10.1016/0022-4073(89)90077-0

Poddar, P., Bandyopadhyay, A., Biswas, D., Ray, B., Ghosh, P.N., 2009. Measurement and analysis of rotational lines in the ($2\nu_1+\nu_2+\nu_3$) overtone band of $H_2O$ perturbed by $CO_2$ using near infrared diode laser spectroscopy. Chem. Phys. Lett. 469, 52–56. doi:DOI: 10.1016/j.cplett.2008.12.074

Pollack, J.B., Dalton, J.B., Grinspoon, D., Wattson, R.B., Freedman, R., Crisp, D., Allen, D.A., Bezard, B., DeBergh, C., Giver, L.P., Ma, Q., Tipping, R.H., 1993. Near-infrared light from Venus' Nightside: A spectroscopic analysis. Icarus 103, 1–42.

Popp, M., Schmidt, H., Marotzke, J., 2016. Transition to a Moist Greenhouse with $CO_2$ and solar forcing. Nat. Commun. 7, 10627. doi:10.1038/ncomms10627

Ramirez, R.M., Kopparapu, R.K., Lindner, V., Kasting, J.F., 2014. Can Increased Atmospheric CO2 Levels Trigger a Runaway Greenhouse? Astrobiology 14, 714–731.

Rothman, L.S., Gordon, I.E., Babikov, Y., Barbe, A., Chris Benner, D., Bernath, P.F., Birk, M., Bizzocchi, L., Boudon, V., Brown, L.R., Campargue, A., Chance, K., Cohen, E.A., Coudert, L.H., Devi, V.M., Drouin, B.J., Fayt, A., Flaud, J.M., Gamache, R.R., Harrison, J.J., Hartmann, J.M., Hill, C., Hodges, J.T., Jacquemart, D., Jolly, A., Lamouroux, J., Le Roy, R.J., Li, G., Long, D.A., Lyulin, O.M., Mackie, C.J., Massie, S.T., Mikhailenko, S., Müller, H.S.P., Naumenko, O. V., Nikitin, A. V., Orphal, J., Perevalov, V., Perrin, A., Polovtseva, E.R., Richard, C., Smith, M.A.H., Starikova, E., Sung, K., Tashkun, S., Tennyson, J., Toon, G.C., Tyuterev, V.G., Wagner, G., 2013. The HITRAN2012 molecular spectroscopic database. J. Quant. Spectrosc. Radiat. Transf. 130, 4–50. doi:10.1016/j.jqsrt.2013.07.002

Rothman, L.S., Gordon, I.E., Barber, R.J., Dothe, H., Gamache, R.R., Goldman, A., Perevalov, V.I., Tashkun, S.A., Tennyson, J., 2010. HITEMP, the high-temperature molecular spectroscopic database. J. Quant. Spectrosc. Radiat. Transf. 111, 2139–2150. doi:10.1016/j.jqsrt.2010.05.001

Sagawa, H., Mendrok, J., Seta, T., Hoshina, H., Baron, P., Suzuki, K., Hosako, I., Otani, C., Hartogh, P., Kasai, Y., 2009. Pressure broadening coefficients of $H_2O$ induced by $CO_2$ for Venus atmosphere. J. Quant. Spectrosc. Radiat. Transf. 110, 2027–2036. doi:10.1016/j.jqsrt.2009.05.003

Segura, T.L., McKay, C.P., Toon, O.B., 2012. An impact-induced, stable, runaway climate on Mars. Icarus 220, 144–148. doi:10.1016/j.icarus.2012.04.013

Segura, T.L., Toon, O.B., Colaprete, A., 2008. Modeling the environmental effects of moderate-sized impacts on Mars. J. Geophys. Res. E Planets 113, 1–15. doi:10.1029/2008JE003147

Segura, T.L., Toon, O.B., Colaprete, A., Zahnle, K., 2002. Environmental Effects of Large Impacts on Mars. Science 298, 1977–1980.

Sung, K., Brown, L.R., Toth, R.A., Crawford, T.J., 2009. Fourier transform infrared spectroscopy measurements of $H_2O$-broadened half-widths of $CO_2$ at 4.3 μm. Can. J. Phys. 87, 469–484. doi:10.1139/P08-068

Tipping, R.H., Boulet, C., Bouanich, J., 1999. for high-temperature $CO_2$ 38, 599–604.

Tran, H., Boulet, C., Stefani, S., Snels, M., Piccioni, G., 2011. Measurements and modelling of high pressure pure $CO_2$ spectra from 750 to 8500cm-1. I-central and wing regions of the allowed vibrational bands. J. Quant. Spectrosc. Radiat. Transf. 112, 925–936. doi:10.1016/j.jqsrt.2010.11.021

Tran, H., Li, G., Ebert, V., Hartmann, J.-M., 2017. Super- and sub-Lorentzian effects in the Ar-broadened line wings of HCl gas. J. Chem. Phys. 146, 194305. doi:10.1063/1.4983397

Tretyakov, M.Y., Serov, E.A., Koshelev, M.A., Parshin, V. V., Krupnov, A.F., 2013. Water dimer rotationally resolved millimeter-wave spectrum observation at room temperature. Phys. Rev. Lett. 110, 1–4. doi:10.1103/PhysRevLett.110.093001

Turbet, M., Forget, F., Svetsov, V., Popova, O., Gillmann, C., Karatekin, O., Wallemacq, Q., Head, J.W., Wordsworth, R., 2017. Catastrophic Events: Possible Solutions to the Early Mars Enigma, in: The Sixth International Workshop on the Mars Atmosphere: Modelling and Observation. Granada, Spain.

Turbet, M., Leconte, J., Selsis, F., Bolmont, E., Forget, F., Ribas, I., Raymond, S.N., Anglada-Escudé, G., 2016. The habitability of Proxima Centauri b II. Possible climates and observability. Astron. Astrophys. 596, A122.

Turbet, M., Tran, H., 2017. Comments on "Radiative transfer in $CO_2$-rich atmospheres: 1. Collisional line mixing implies a colder early Mars." J. Geophys. Res. accepted.

Turbet, M., Tran, H., Hartmann, J.-M., Forget, F., 2017. Toward a more accurate Spectroscopy of CO2/H2O-Rich Atmospheres: Implications for the Early Martian Atmosphere, in: Fourth International Conference on Early Mars: Geologic, Hydrologic, and Climatic Evolution and the Implications for Life, Proceedings of the Conference Held 2-6 October, 2017 in Flagstaff, Arizona. LPI Contribution No. 2014, 2017, id.3063.

Wordsworth, R. D., Pierrehumbert, R. T., 2013. Water Loss from Terrestrial Planets with CO2-rich Atmospheres. The Astrophysical Journal, Vol. 778, Issue 2, article id. 154, 19 pp. doi:10.1088/0004-637X/778/2/154